\documentclass[aps,prl,reprint,superscriptaddress,showpacs]{revtex4-1}
\usepackage{graphicx}
\usepackage{epsfig}
\usepackage{amsmath,amsthm}
\usepackage{textcomp}
\usepackage{color}
\usepackage{multirow}
\usepackage{pdfpages}

\begin{document}

\title{Evidence for eight node mixed-symmetry superconductivity in a correlated organic metal}
\author{Daniel Guterding}
\affiliation{Institut f{\"u}r Theoretische Physik, Goethe-Universit\"at Frankfurt, Max-von-Laue-Str. 1, 60438 Frankfurt am Main, Germany}
\author{Sandra Diehl}
\affiliation{Graduate School Materials Science in Mainz, Staudingerweg 9, 55128 Mainz, Germany}
\affiliation{Institut f\"ur Physik, Johannes Gutenberg-Universit\"at Mainz, Staudingerweg 7, 55128 Mainz, Germany}
\author{Michaela Altmeyer}
\affiliation{Institut f{\"u}r Theoretische Physik, Goethe-Universit\"at Frankfurt, Max-von-Laue-Str. 1, 60438 Frankfurt am Main, Germany}
\author{Torsten Methfessel}
\affiliation{Institut f\"ur Physik, Johannes Gutenberg-Universit\"at Mainz, Staudingerweg 7, 55128 Mainz, Germany}
\author{Ulrich Tutsch}
\author{Harald Schubert}
\author{Michael Lang}
\author{Jens M\"uller}
\author{Michael Huth}
\affiliation{Physikalisches Institut, Goethe-Universit\"at Frankfurt, Max-von-Laue-Str. 1, 60438 Frankfurt am Main, Germany}
\author{Harald O. Jeschke}
\author{Roser Valent\'i}
\affiliation{Institut f{\"u}r Theoretische Physik, Goethe-Universit\"at Frankfurt, Max-von-Laue-Str. 1, 60438 Frankfurt am Main, Germany}
\author{Martin Jourdan}
\author{Hans-Joachim Elmers}
\email{elmers@uni-mainz.de}
\affiliation{Institut f\"ur Physik, Johannes Gutenberg-Universit\"at Mainz, Staudingerweg 7, 55128 Mainz, Germany}

\date{\today}

\begin{abstract}
We report a combined theoretical and experimental investigation of the 
superconducting state in the quasi-two-dimensional organic superconductor 
$\kappa$-(ET)$_2$Cu[N(CN)$_2$]Br. Applying spin-fluctuation theory to a 
low-energy material-specific Hamiltonian derived from {\it ab initio} density 
functional theory we calculate the quasiparticle density of states in the 
superconducting state. We find a distinct three-peak structure that results from 
a strongly anisotropic  mixed-symmetry superconducting gap with eight nodes and  
twofold rotational symmetry. This theoretical prediction is supported by 
low-temperature scanning tunneling spectroscopy on {\it in situ} cleaved single 
crystals of $\kappa$-(ET)$_2$Cu[N(CN)$_2$]Br with the tunneling direction 
parallel to the layered structure. 
\end{abstract}


\maketitle
%
%
{\it Introduction.- }
It has been widely accepted that $\kappa$-(ET)$_2 X$ organic
charge-transfer salts [where ET denotes
bis(ethylenedithio)tetrathiafulvalene and $X$ a monovalent anion] share essential
features with cuprates regarding their superconducting
state~\cite{McKenzie1997}. In both classes of materials the electronic structure
is quasi-two-dimensional and superconductivity emerges in the vicinity of an
antiferromagnetically ordered Mott insulating phase. The transformation of these
Mott insulators into superconductors can be achieved either by doping (cuprates)
or by increasing the bandwidth (organics) through the application of physical or
chemically-induced pressure, see e.g.\ Ref.~\onlinecite{Toyota2007}. In view of
these similarities it is natural to investigate whether a related coupling
mechanism is operative in both classes of materials.

A direct determination of the pairing mechanism is extremely difficult, if not 
impossible, and therefore most efforts have been directed towards the 
determination of the symmetry of the superconducting energy gap function. For 
the cuprates, thanks to the availability of phase-sensitive probes, compelling 
evidence was provided early on that the gap has predominantly $d_{x^2-y^2}$ 
symmetry~\cite{vanHarlingen1995,Kirtley2000}. The lack of such probes for the 
organic charge-transfer salts, rooted in difficulties in proper material 
preparation, makes the situation less clear. There has been a long-lasting 
controversy about the gap symmetry in the most widely studied materials 
$\kappa$-(ET)$_2$Cu[N(CN)$_2$]Br (in short $\kappa$-Br) and 
$\kappa$-(ET)$_2$Cu(NCS)$_2$ ($\kappa$-NCS) with some results in support of an 
$s$-wave and others consistent with a $d$-wave scenario (see the 
reviews~\cite{LangMueller2004,Kuroki2006,PowellMcKenzie2006,Toyota2007} for the 
status of the discussion up to the year 2006). More recent attempts include 
magnetocalorimetry~\cite{Malone2010}, surface impedance~\cite{Milbradt2013} and 
high-resolution specific heat measurements~\cite{Taylor2007,Taylor2008}, both in 
favor of $d$-wave pairing, even though earlier results of the specific heat 
favored $s$-wave symmetry~\cite{Elsinger2000,Mueller2002,Wosnitza2003}. On the 
other hand, by analyzing measurements of elastic constants~\cite{Dion2009} a 
mixed order parameter of either $A_{1g}+ B_{1g}$ or $B_{2g}+B_{3g}$ has been 
claimed for $\kappa$-Br following  the classification of irreducible 
representations of the material's orthorhombic $D_{2h}$ point group.

In the above-mentioned studies the gap function could be determined only
indirectly from the temperature dependence of the quantity investigated. In
contrast, a direct examination of the energy gap and its angular dependence is
possible, in principle, through scanning tunneling spectroscopy (STS). Such STS
studies have been performed by tunneling either into {\it as
grown} surfaces for $\kappa$-NCS~\cite{Arai2001} and
$\kappa$-Br~\cite{Ichimura2008}, or into {\it focused-ion-beam-cut} surfaces for
partially deuterated $\kappa$-Br~\cite{Oka2015}. The results were found to be
consistent with a $d$-wave order parameter. However, with the exception of
Ref.~\onlinecite{Oka2015}, no surface characterization was provided in these
studies.

In this Letter we report evidence for an anisotropic superconducting gap 
structure for $\kappa$-Br based on a combined theoretical and STS investigation. 
By first constructing a two-dimensional Hubbard model of the ET
layer from {\it ab initio} density functional theory (DFT) calculations and 
subsequently applying spin-fluctuation theory to this model, we predict the 
presence of a  strongly anisotropic superconducting gap. As a consequence of the 
orthorhombicity of the crystal structure and the resulting details of the 
electronic structure the order parameter does not have pure $d$-wave symmetry. 
Instead, an eight node twofold rotationally symmetric state is realized, which 
we identify as (extended) $s+d_{x^2-y^2}$. This pairing symmetry results in a 
quasiparticle density of states (DOS) with a distinct three-peak structure. STS 
measurements performed  on well-defined surfaces obtained from a recently 
developed {\it in situ} cleaving technique~\cite{Diehl2014} support this 
theoretical finding. 

%
%
{\it Theory.- }
We performed {\it ab initio} density functional theory (DFT) calculations for 
$\kappa$-Br within an all-electron full-potential local orbital 
(FPLO)~\cite{FPLOmethod} basis and subsequently derived  a tight-binding model 
from projective molecular orbital Wannier functions~\cite{FPLOtightbinding, 
EndgroupDisorder}. The resulting low-energy Hamiltonian $H_0 =\sum_{ij} 
t_{ij}(c^\dagger_{i} c^{\,}_{j} + h.c.)$, with hopping parameters $t_{ij}$ 
between ET molecules at sites $i$ and $j$,   consists of four bands arising from 
the highest occupied molecular orbitals of the ET molecules in the 
crystallographic unit cell and is 3/4-filled~(see 
Ref.~\onlinecite{EndgroupDisorder}). 
   
\begin{figure}[tb]
\includegraphics[width=0.8\linewidth]{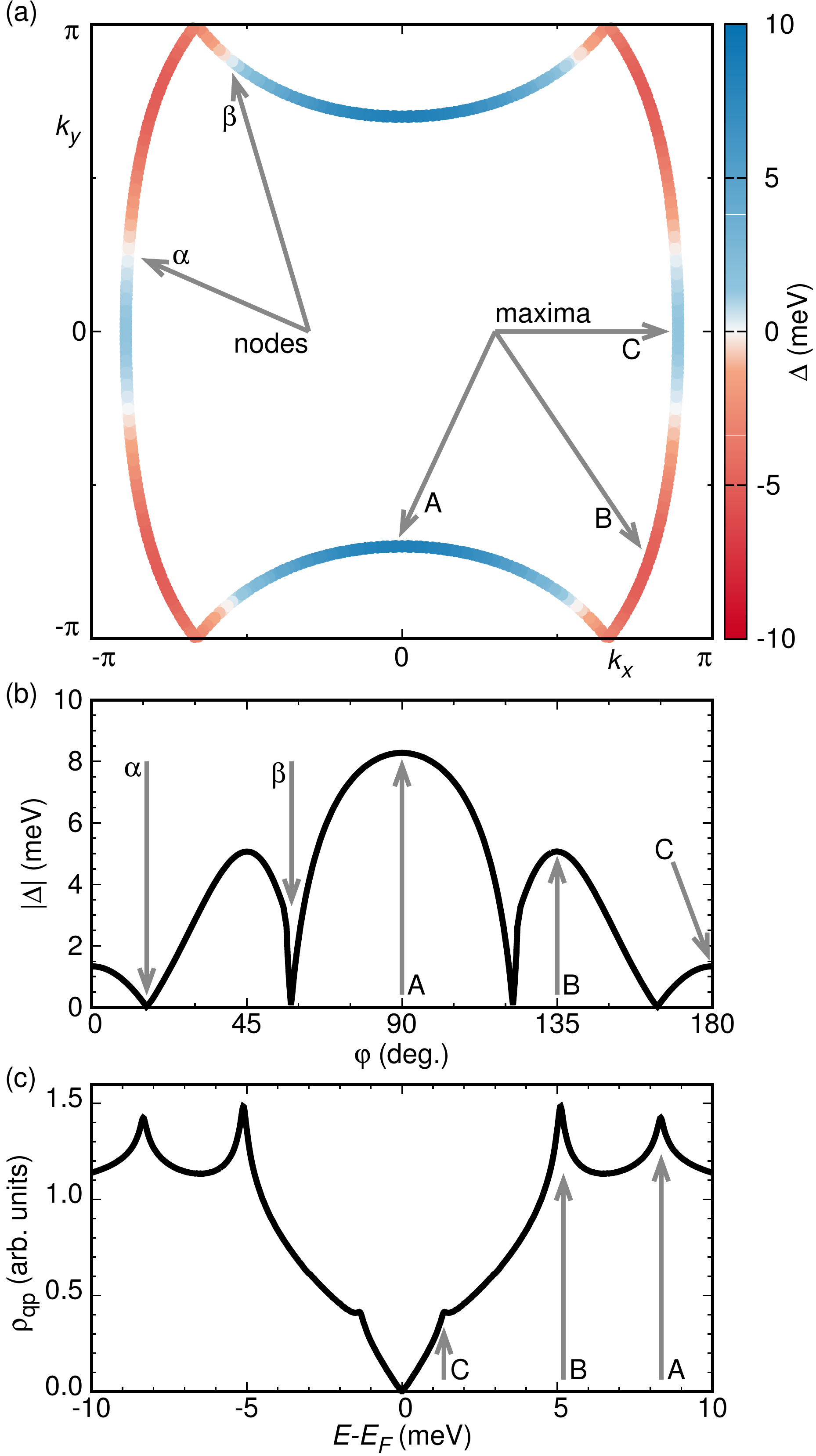}
\caption{\label{fig:theory}(Color online) {\it Ab initio} calculated results for 
(a) superconducting gap $\Delta$ on the Fermi surface, (b) absolute value $| 
\Delta |$ of the superconducting gap as a function of the angle $\varphi$ in the 
$k_x$-$k_y$-plane measured with respect to the $k_x$-direction and (c) 
quasiparticle DOS $\rho_{qp}$ in the superconducting state of $\kappa$-Br. The 
maxima of the superconducting gap magnitude are indicated by arrows A, B and C, 
while the positions of nodes are indicated by arrows $\alpha$ and $\beta.$ The 
energy scale is set to $\Delta_0 = 10~\mathrm{meV}$. A small broadening of 
$\Gamma = 0.07~\mathrm{meV}$ is included, here.}
\end{figure}

Previous theoretical approaches~\cite{Schmalian1998} approximated the real 
crystal structure by an anisotropic triangular lattice of (ET)$_2$ 
dimers~\cite{HoppingStructureDifference}. This approximation was shown to be 
justified in the insulating phase of $\kappa$-(ET)$_2 X$ 
materials~\cite{Kino1996}, but not in the superconducting 
state~\cite{Kuroki2002}. Hence our method, which is based on an {\it ab 
initio} derived Hamiltonian with the full symmetry of the ET
layer, is more realistic than previous approaches.

In the following we take the rigorous viewpoint that superconductivity in 
$\kappa$-Br is solely driven by electron-electron interactions, see 
e.g.~\cite{Toyota2007,WosnitzaReview,PowellMcKenzieSpinFluctuationsNMR}. 
Therefore, we add an intramolecular Hubbard interaction term $H_{int} = U\sum_i 
n_{i \uparrow} n_{i \downarrow}$ to $H_0$. In this setup we calculate the 
optimal geometry of the superconducting gap function using random phase 
approximation (RPA) spin-fluctuation theory in the spin-singlet 
channel~\cite{NJPSpinfluctuationMethod}. For computational details see 
Ref.~\onlinecite{Supplement}. We stress, however, that our results do not 
exclude a contribution from electron-phonon interactions, indications of which 
have been found, e.g., in studies of the isotope effect~\cite{Kini1997} and 
phonon renormalization~\cite{Pintschovius1997}.

From the RPA calculation we find that the superconducting order parameter has 
eight nodes [see Fig.~\ref{fig:theory}(a)] in contrast to previous reports of 
pure $d$-wave states with four nodes~\cite{Schmalian1998, Kuroki2006}. The first 
set of nodes labelled $\alpha$ in Fig.~1(a) appears on the quasi-one-dimensional 
part of the Fermi surface. The four nodes on the elliptic part of the Fermi 
surface, denoted $\beta$ in Fig.~\ref{fig:theory}~(a), are close to the 
Brillouin zone diagonals. The full angular dependence of the gap function is 
shown in Fig.~\ref{fig:theory}(b). The angles corresponding to the node positions 
are $\varphi_\alpha \sim 15^\circ$ and $\varphi_\beta \sim 58^\circ$, all 
measured with respect to the $k_x$-direction. Nodes at other angles are symmetry 
equivalent to $\alpha$ and $\beta$. The three maxima of the superconducting gap 
magnitude are indicated by letters A, B and C in Fig.~\ref{fig:theory}. The 
corresponding angles are $\varphi_A = 90^\circ$, $\varphi_B = 45^\circ$ and 
$\varphi_C = 0^\circ$. Maxima at other angles are symmetry equivalent to A, B 
and C.

The quasiparticle DOS $\rho_{qp}(E)$ in the superconducting state can be 
calculated from the momentum structure of the superconducting gap $\Delta(\vec 
k)$, as given in Eq.~\ref{eq:dosqpfinal}. A small broadening of the energy 
spectrum due to finite quasiparticle lifetimes is modeled by the parameter 
$\Gamma$~\cite{Dynes1978}.
\begin{equation}
\rho_{qp}(E, \Gamma) \propto \sum\limits_{\vec k} 
\,\textnormal{Re}\frac{ \vert E+i\Gamma \vert}
{\sqrt{(E+i\Gamma)^2-\Delta(\vec k)^2 }}
\label{eq:dosqpfinal}
\end{equation}
The main difference between our formulation (Eq.~\ref{eq:dosqpfinal}) and 
conventional approaches~\cite{Arai2001, Ichimura2008, Oka2015} is that we do 
{\it not} assume the Fermi surface to be a concentric circle. Instead we 
transform the usual angular integration (see 
Refs.~\onlinecite{Supplement,Arai2001, Ichimura2008, Oka2015}) into a sum over 
the discretized {\it ab initio} Fermi surface. Our approach  can therefore 
discriminate between different $d$-wave solutions already on the level of the 
quasiparticle DOS. The approximations made in the derivation of 
Eq.~\ref{eq:dosqpfinal} are detailed in Ref.~\onlinecite{Supplement}. The 
quasiparticle DOS calculated from Eq.~\ref{eq:dosqpfinal}, using the gap 
calculated from RPA for $\Delta (\vec k)$, shows three distinct features 
[Fig.~\ref{fig:theory}(c)] that correspond to the gap maxima A, B and C already 
discussed.

%
%
{\it Experiment.- }
Single crystals of $\kappa$-Br were grown in an electrochemical crystallization 
process~\cite{Anzai1995}. All investigations were performed with a commercial 
low-temperature scanning tunneling microscope (STM) under ultra-high vacuum 
(UHV) conditions with a base pressure of about $5\cdot10^{-11}$\,mbar. In order 
to obtain a clean surface as is essential for high-resolution STM 
investigations, we cleaved the crystal surface with a homebuilt cutter in UHV. 
After {\it in situ} cleaving, the sample was mounted into the pre-cooled STM 
stage. Measurements have been performed on three single crystals, yielding 
essentially identical results. 

By this procedure, the cooling rate of the sample, relevant at the glass-like 
ethylene-endgroup ordering transition $T_g$,  is estimated to be about $- 
1$\,K/min at $T_g \approx 75-80$\,K (see Ref.~\onlinecite{Muller2002}). This 
implies a residual disorder in the orientational degrees of freedom of the ET 
molecules' terminal ethylene groups of order $3$~\% 
~\cite{Hartmann2014,Mueller2015}. For details on the STM/STS measurements, see 
Ref.~\onlinecite{Diehl2014}.

\begin{figure}[tb]
\includegraphics[width=0.8\columnwidth]{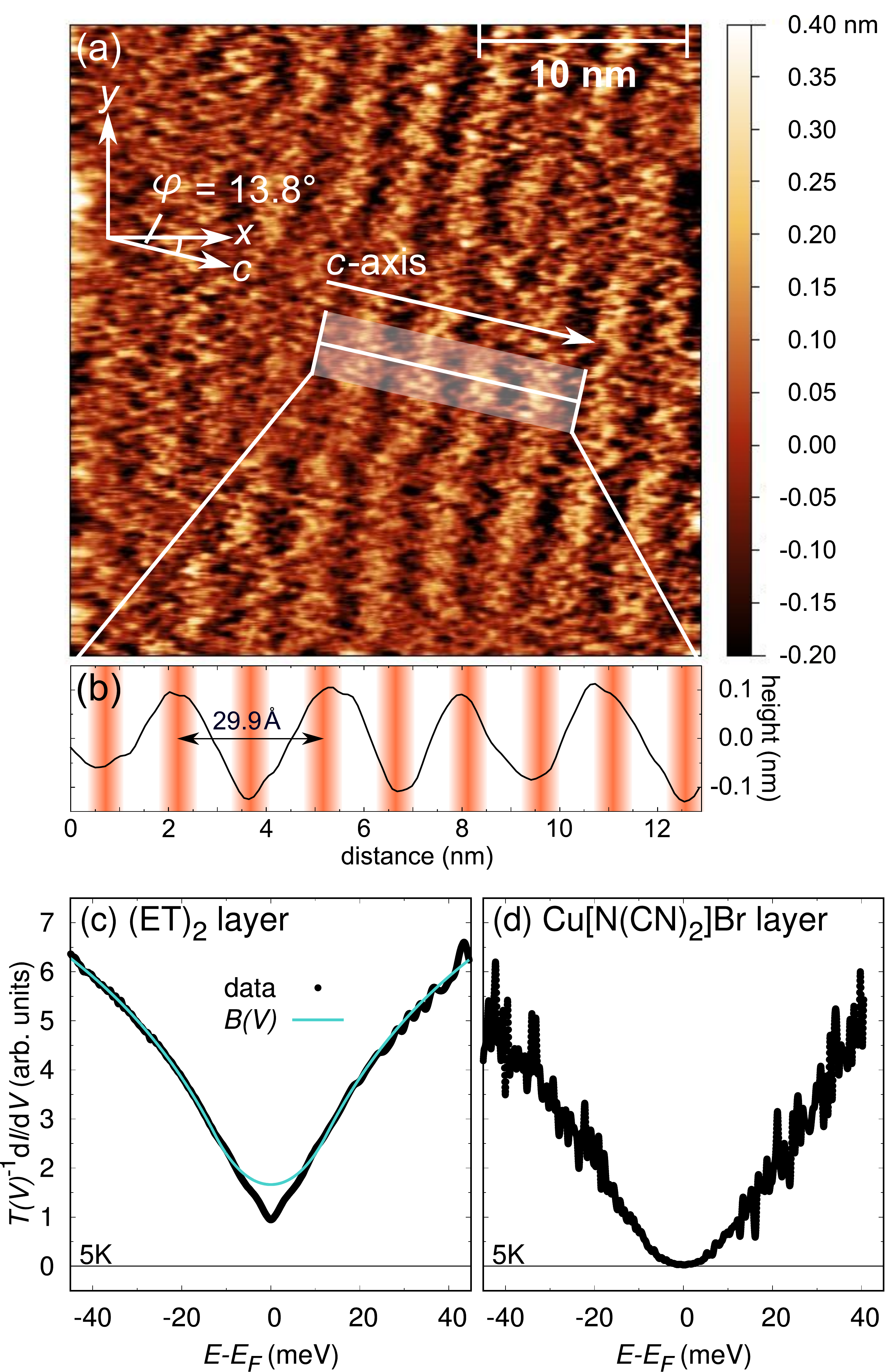}
\caption{\label{fig:exp}(Color online) (a) (30\,x\,30)\,nm$^2$ STM image of the 
crystal surface with tunneling direction parallel to the  layered structure of 
the $\kappa$-Br crystal revealing a stripe pattern ($T$\,=\,5\,K, 
$I$\,=\,60\,pA, $U_{\rm tip} = 30$\,mV). (b) The height profile along the marked 
area in (a) reveals an average stripe width of $29.9$\,\AA. The red stripes 
indicate the position of the insulating layers. (c) Conductance spectra  
d$I$/d$V / T(V)$ at $T = 5~\mathrm{K}$ (black dots) measured at the ET layer and 
(d) the Cu[N(CN)$_2$]Br layer with $-eU_{\rm tip}=eV=E-E_F$. The function $B(V)$ 
is also plotted in (c) (see text for discussion).} 
\end{figure}

Fig.~\ref{fig:exp}(a) shows a (30\,x\,30)\,nm$^2$ STM image of a $\kappa$-Br 
crystal cleaved perpendicular to the molecular layers (for the parallel 
topography see Refs.~\onlinecite{Diehl2014,Burgin1995}). The STM image reveals a 
stripe pattern with an average width of 29.9\,\AA\ [see Fig.~\ref{fig:exp}(b)] 
which is in good agreement with the lattice constant of $\kappa$-Br in the 
$c$-direction. One period of the stripe pattern consists of two ET and two anion 
layers. For a better illustration Fig.~\ref{fig:exp}(b) shows the height profile 
along the white shaded area in (a), wherein the position of the anion layers is 
marked by red stripes. The height profile is mainly a mapping of the 
topographical surface profile, but is also influenced by the local density of 
states. By analyzing the height profile along a single unit cell we conclude 
that the cutting plane is rotated by $\varphi_{ab}=60$\textdegree\ about the 
$c$-axis with respect to the $b$-axis (see Ref.~\onlinecite{Supplement}).

Next, we measured the differential conductance d$I$/d$V$ of the ET layer and the 
anion layer in the temperature range from 5\,K to 13\,K [see 
Fig.~\ref{fig:exp}(c) and (d) for the data at 5\,K] in order to extract 
information about their DOS. The d$I$/d$V$ spectra have been 
normalized to the tunneling transmission function $T(V)$, which is almost linear 
in the measured energy range and accounts for a small asymmetry observed in the 
original spectra~\cite{Ukraintsev1996}. Assuming a constant DOS for the tip 
material, (d$I$/d$V$)/$T(V)$ is proportional to the differential conductance 
$D(V)$ of the sample.

The differential conductance is non-zero at $E=E_\mathrm{F}$ for the spectrum 
measured parallel to the ET layer [Fig.~\ref{fig:exp}(c)]. In contrast, the 
differential conductance spectrum measured at the anion layer 
[Fig.~\ref{fig:exp}(d)] approaches zero at the Fermi edge [d$I$/d$V 
(E=E_F)\approx 0$] indicating the expected insulating behavior. The $V$-shaped 
feature for $|eV|>15$~meV results from a logarithmic suppression of the density 
of states at the Fermi edge described by a function $B(V)$ [see 
Fig.~\ref{fig:exp}(c)] which has been assigned to electronic 
disorder~\cite{Diehl2014}. Temperature-dependent measurements up to 13\,K (see 
Ref.~\onlinecite{Supplement}) show a nearly temperature-independent function for 
the DOS $B(V)$ as described by Refs.~\onlinecite{Shinaoka2009,Diehl2014}.

\begin{figure}[t]
\includegraphics[width=1\columnwidth]{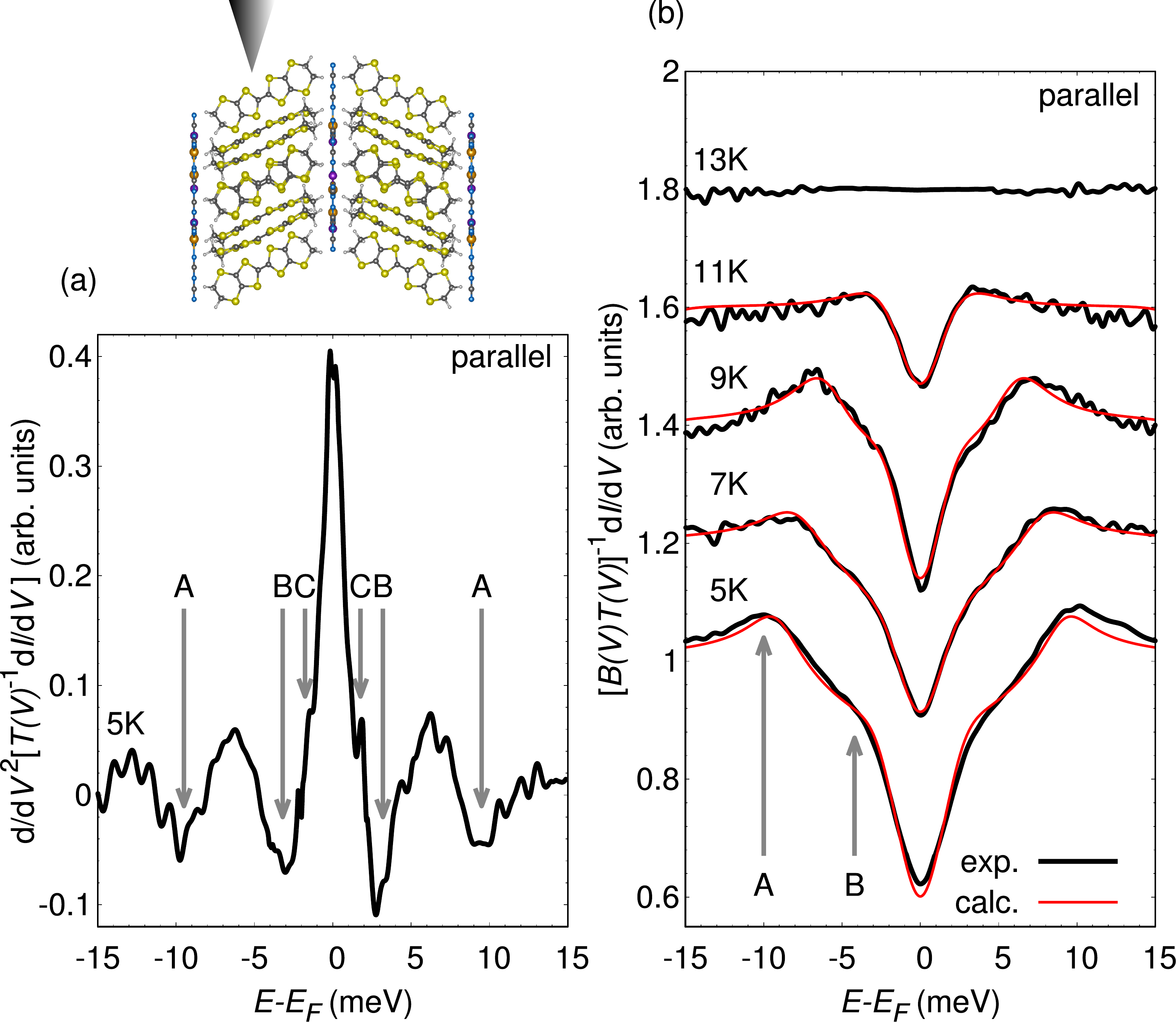}
\caption{\label{fig:comb}(Color online) (a) Second derivative of the conductance 
spectrum d$I$/d$V / T(V)$ at $T = 5~\mathrm{K}$ shown in Fig.~\ref{fig:exp}(c). 
Arrows labeled by A, B and C indicate three pairs of minima which are symmetric 
to the origin, corresponding to the coherence peaks partially seen in (b).   (b) 
Conductance spectra $S(V) = 1/[B(V)T(V)]\;dI/dV$ of the superconducting state as 
a function of $eV=E-E_F$ at different temperatures measured parallel to the 
layered crystal structure. The red lines show mappings of 
Eq.~(\ref{eq:fitformula}) to the measured data.}
\end{figure}

An important piece of information is obtained from the second derivative of 
d$I$/d$V$ w.r.t. the voltage as presented in Fig.~\ref{fig:comb}(a). This 
function at $T = 5~\mathrm{K}$ shows three pairs of narrow minima (denoted as A, 
B and C), which correspond to superconducting coherence peaks and will be 
discussed further below. The pronounced maximum at $E_F$ reflects the zero-bias 
anomaly of the normal-state DOS~\cite{Diehl2014}.

The conductance spectrum related to the superconducting phase is given by $S(V) 
= [B(V)T(V)]^{-1}\;dI/dV$. In Fig.~\ref{fig:comb}(b), $S(V)$ measured parallel 
to the layered crystal structure is plotted for temperatures between 5\,K and 
13\,K. The two main features in the second derivative [shown by arrows A and B 
in Fig.~\ref{fig:comb}(b)] are already visible in the bare conductance spectra, 
whereas feature C becomes discernible only in the second derivative of the spectra. 

%
%
{\it Discussion.-}
Our theoretical calculations predict an anisotropic superconducting order 
parameter. We note that a very accurate representation of the calculated gap on 
the Fermi surface is given in terms of extended $s$- and $d$-wave functions (see 
also Ref.~\onlinecite{Supplement}): $\Delta (\vec k) = \Delta_0 [ c_{s_1} 
(\text{cos}~k_x + \text{cos}~k_y) + c_{d_1} (\text{cos}~k_x - \text{cos}~k_y) + 
c_{s_2} (\text{cos}~k_x \cdot \text{cos}~k_y)]$. Although our experiments 
resolve the existence of feature C only in the second derivative of the 
conductance, the shape of the experimental tunneling spectra 
[Fig.~\ref{fig:comb}(b)] bears the essential features predicted by our theory 
[Fig.~\ref{fig:theory}(c)]. Features A and B are only slightly shifted with 
respect to each other.

In order to connect the calculated quasiparticle DOS (Eq.~\ref{eq:dosqpfinal}) 
to the measured spectrum $S(V)$  we investigate whether quantitative agreement 
can be reached by fine-tuning some parameters. In this process the symmetry of 
the superconducting state is kept fixed to the theoretical prediction (mixed 
extended $s+d_{x^2-y^2}$). We calculate $S(V)$ from Eq.~\ref{eq:fitformula} (see 
Ref.~\onlinecite{Supplement}), where $f(E)$ is the Fermi function, $V$ is the 
bias voltage and $x$ is a background shift, which accounts for parasitic 
conduction paths.
\begin{equation}
\begin{array}{rl}
S(V) =& \frac{1}{B(V)T(V)}\frac{dI(V)}{dV}\\[4pt]
\propto &  \int_{-\infty}^\infty dE \,
\left[ \rho_{qp}(E) (1-x) +x \right]
\frac{-df(E+eV)}{dV}
\end{array}
\label{eq:fitformula}
\end{equation}

Using the expression for $\Delta (\vec k)$ given above, we re-evaluate 
Eqs.~\ref{eq:dosqpfinal} and \ref{eq:fitformula} with parameter sets 
$\{\Delta_0$, $c_{s_1}$, $c_{d_1}$, $c_{s_2}$, $x$, $\Gamma\}$ until optimal 
agreement with the experimental spectra in the interval $[-12, 
+12]~\mathrm{meV}$ is reached. The corresponding calculated spectra are shown in 
Fig.~\ref{fig:comb}(b) as solid red lines. The optimal parameter values listed in 
Table~\ref{tab:fitparameters} are consistent throughout the investigated 
parameter range. Only the origin of the non-monotonous behavior of $\Gamma$ is 
currently unclear. 

\begin{table}[t]
\caption{Values of the parameters in Eq.~\ref{eq:fitformula}
obtained by mapping the calculated $S(V)$ to the experimental spectra.}
\label{tab:fitparameters}
\begin{ruledtabular}
\begin{tabular}{ccccccc}
$T$ (K) & $c_{s_1}$  & $c_{d_1}$ & $c_{s_2}$
&$\Delta_0$ (meV) & $\Gamma$ (meV) & x\\
\hline
 5 &-0.109 &-0.276 &-0.615 &12.218& 0.690& 0.520\\
 7 &-0.128 &-0.280 &-0.592 &10.638& 0.641& 0.603\\
 9 &-0.064 &-0.317 &-0.620 & 7.376& 0.000& 0.466\\
11 &-0.158 &-0.200 &-0.642 & 2.984& 0.035& 0.188\\
\end{tabular}
\end{ruledtabular}
\end{table}

The nodal positions we find in our study are particularly appealing because of the 
ongoing controversy about the realization of either $d_{xy}$ and $d_{x^2-y^2}$ 
pairing in quasi-2D organic superconductors. In previous works, a fourfold 
rotational symmetry constraint was introduced based on  single-band Hubbard-type 
models  proposed for these systems with a dimer of molecules per site on an 
anisotropic triangular 
lattice~\cite{Kino1996,McKenzie1998,Schmalian1998,Zehetmayer2013,Kandpal2009}. 
Inspired by recent {\it ab initio} studies~\cite{Nakamura2012,Altmeyer2015} we 
however went back to the original $\kappa$-type lattice structure and treated all 
molecules as entities. As a consequence,  we find evidence for a mixed-symmetry 
extended $s+d_{x^2-y^2}$ state, which has been suggested to exist in some 
parameter regions of various models for $\kappa$-type charge-transfer 
salts~\cite{Kuroki2002, PowellMcKenzieHubbardHeisenberg, Watanabe2008}, but has 
so far not been shown to exist in a material-specific Hamiltonian. 
Interestingly, a different kind of eight node superconductivity has been 
recently proposed in the context of Fe-based superconductors~\cite{Okazaki}.

The nodes $\beta$ correspond to the usually discussed $d_{x^2-y^2}$ solution, 
while the nodes $\alpha$ lie close to the $k_x$-direction. In contrast to the 
$d_{xy}$-gap found in Ref.~\onlinecite{Schmalian1998} there are no nodes along 
the $k_y$-direction in our results. The twofold rotational symmetry found in 
our study is very important, because the analysis of many experiments on 
$\kappa$-(ET)$_2 X$ materials is based, possibly mistakenly, on the subtraction 
of a large twofold symmetric alleged background (see f.i. 
Refs.~\onlinecite{Izawa2002, Malone2010}) attributed to other effects, such as 
phonons.

We note that from a study of the suppression of $T_c$ with increasing disorder 
in $\kappa$-NCS, a mixed order parameter has been suggested as an important 
constraint on models of the superconductivity \cite{Analytis2006}. The 
anisotropic gap should also appear in related physical properties, such as the 
electronic part of the specific heat. Calculations of the electronic 
contribution to the specific heat based on the gap function determined in this 
work are in favorable agreement with experimental observations, see 
Ref.~\onlinecite{Supplement}.

%
%
{\it Summary.-}
We have studied the organic superconductor $\kappa$-(ET)$_2$Cu[N(CN)$_2$]Br in 
RPA spin-fluctuation theory based on {\it ab initio} density functional theory 
calculations, as well as low-temperature scanning tunneling microscopy and 
spectroscopy with the tunneling direction parallel to the layered structure. In 
both theory and experiment we find evidence for three coherence peaks in the 
spectra of the superconducting state. Our results indicate that the symmetry of 
the superconducting pairing in $\kappa$-(ET)$_2 X$ organics might be neither of 
the $d_{xy}$ and $d_{x^2-y^2}$ states discussed in previous studies, but of a 
mixed extended $s+d_{x^2-y^2}$ type with four nodes  located close to the 
Brillouin zone diagonals and four nodes located on the quasi-1D sheets. 

\begin{acknowledgments}
{\it Acknowledgments.-}
We thank the German Research Foundation (Deutsche Forschungsgemeinschaft, SFB/TR 
49) and the Graduate School Materials Science in Mainz for financial support. DG 
acknowledges helpful discussions with Andreas Kreisel, Peter J. Hirschfeld and 
Paul C. Canfield. Calculations were performed on the LOEWE-CSC and FUCHS 
supercomputers of the Center for Scientific Computing (CSC) in Frankfurt am 
Main, Germany.
\end{acknowledgments}

\bibliographystyle{apsrev4-1}

\clearpage


\section*{Supplementary material}

# Evidence for eight node mixed-symmetry superconductivity in a correlated organic metal: Supplemental Information

Daniel Guterding,[1] Sandra Diehl,[2,3] Michaela Altmeyer,[1] Torsten Methfessel,[3]
Ulrich Tutsch,[4] Harald Schubert,[4] Michael Lang,[4] Jens Müller,[4] Michael Huth,[4]
Harald O. Jeschke,[1] Roser Valentí,[1] Martin Jourdan,[3] and Hans-Joachim Elmers[3, *]

[1]Institut für Theoretische Physik, Goethe-Universität Frankfurt,
Max-von-Laue-Str. 1, 60438 Frankfurt am Main, Germany
[2]Graduate School Materials Science in Mainz, Staudingerweg 9, 55128 Mainz, Germany
[3]Institut für Physik, Johannes Gutenberg-Universität Mainz, Staudingerweg 7, 55128 Mainz, Germany
[4]Physikalisches Institut, Goethe-Universität Frankfurt,
Max-von-Laue-Str. 1, 60438 Frankfurt am Main, Germany
(Dated: May 18, 2016)

## I. EXPERIMENTAL CRYSTAL ORIENTATION

The height profile measured across the insulating and conducting layers shown in the STM image (Fig. 1 in the main text) carries mostly topographical information. The insulating anion-layers are expected to appear slightly lower (darker) than the conducting BEDT-TTF layers, which is indeed observed but it is, however, a small contribution. Therefore, one can deduce the tilting angle from a comparison to the known crystal structure. By analyzing the height profile along a single unit cell we determine the orientation of the cutting plane (see Fig. 1). Depending on the rotation about the $c$-axis (the axis perpendicular to the conducting layers) the height corrugation caused by the alternating orientation of the BEDT-TTF molecules is more or less pronounced. The best agreement between height profile and crystal structure is found for a tilting angle $\varphi_{ab} = 60°$ about the $c$-axis with respect to the $b$-axis.

## II. EXPERIMENTAL EXTRACTION OF THE SUPERCONDUCTING DOS

In STS, the spectra have to be corrected for the different work functions of tip and sample which leads to a voltage-dependent asymmetric tunneling transmission function $T(V)$[1]. For this reason all spectra shown here are normalized to $T(V)$ which was obtained from spectra taken at 13 K, $i.e.$, in the normal state of $\kappa$-Br. We furthermore assume a constant DOS for the tip material in the relevant energy range so that the $dI/dV/T(V)$ curves reflect the thermally smeared DOS $D(V)$ of the sample with $eV = E - E_F$ ($E_F$ denotes the Fermi energy). Then we analyze this data in the framework of the Anderson-Hubbard (AH) model discussed by Shinaoka and Imada[2] for disordered itinerant electron systems with short-range interactions[3].

In their numerical treatment of the AH model a scaling law is introduced for the DOS in the presence of short-range Coulomb interactions and a multi-valley energy landscape[2]: $B(V) = c \exp[-\alpha \left(-\log |eV|\right)^d]$, $|V| \geq V_0$,

where $d$ denotes the spatial dimension ($d = 2$ for the present case) and $\alpha = 0.288$ is a non-universal constant. In the immediate vicinity of $E_F$ the AH model is not applicable[2]. A very good description of this energy region is obtained by assuming a hard energy gap of small size accounted for by a phenomenological DOS function of the following form: $B(V) = c' \cosh(\frac{eV}{\varepsilon_T})$, $|V| < V_0$, where $c'$ is a constant and $\varepsilon_T$ measures the effective barrier height. For finite temperatures $T > 0$, $B(V)$ rapidly becomes non-zero near $E_F$ because a thermally activated crossing of the small barrier $\varepsilon_T$ leads to a nearly temperature-independent prefactor and the voltage-dependence is described by the cosh term. With the requirement of continuous differentiability at the inflection point $V_0$, $c'$ and $\varepsilon_T$ terms are fixed for any given temperature and do not represent adjustable parameters. The conductance spectrum of the superconducting state is given by

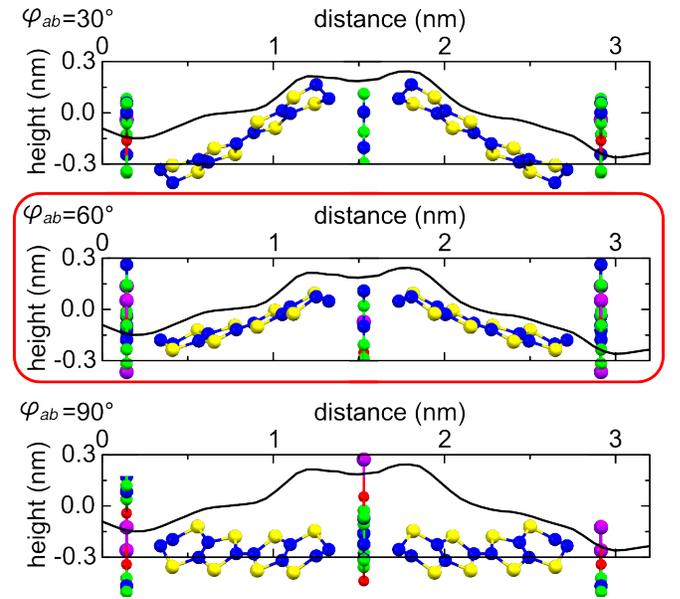

FIG. 1. Height profile along a single unit cell with the corresponding crystal orientation showing the best agreement for an angle for rotation about the $c$-axis of $\varphi_{ab} = 60°$.

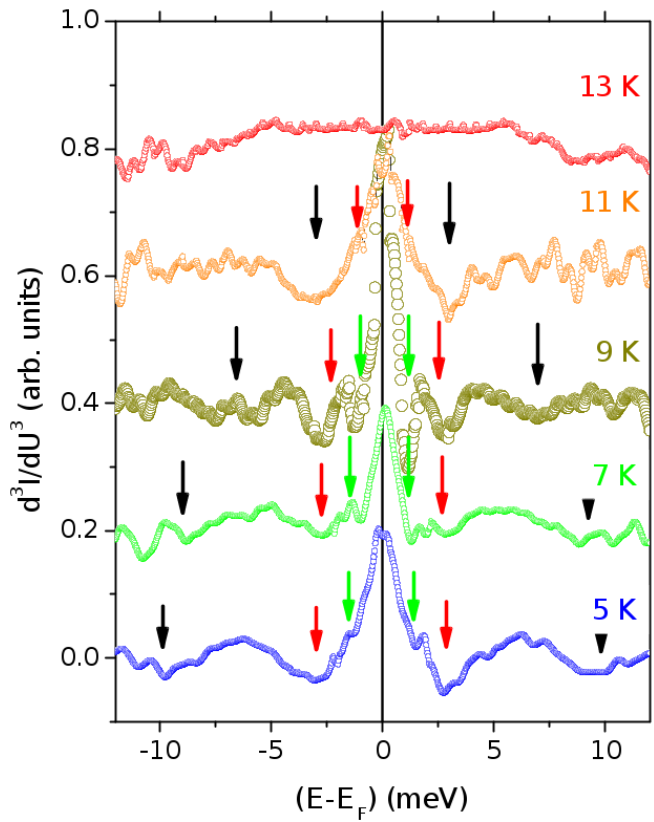

FIG. 2. Temperature dependence of the second derivative of the conductance spectrum $\mathrm{d}I/\mathrm{d}V/T(V)$. Features A, B and C are indicated by the colored arrows (black, red and green respectively).

$$S(V) = dI/dV/T(V)/B(V).$$

## III. TEMPERATURE DEPENDENCE OF FEATURES OBSERVED IN THE RAW CONDUCTANCE

In Fig. 3 of the main paper we have shown the second derivative of the conductance spectrum $\mathrm{d}I/\mathrm{d}V/T(V)$ at $T = 5$ K. For completeness, the temperature dependence of this quantity is shown in Fig. 2. The temperature evolution of the superconducting gaps associated with features A, B and C can be traced clearly. All gaps decrease when increasing the temperature towards $T_c$. In the spectrum taken at $T = 13$ K $> T_c$ all signatures of a superconducting gap are gone.

The second derivative of the tunneling conductance was calculated after smoothing the data using the Savitzky-Golay method[4], which is a widely used technique to increase the signal-to-noise ratio without significantly distorting the signal data. The averaging interval is 2.3 meV in the interval [-2.09:2.09] meV and 6 meV outside of this interval. Different smoothing intervals are justified because the noise increases with increasing absolute tunneling current. Thus the noise is considerably

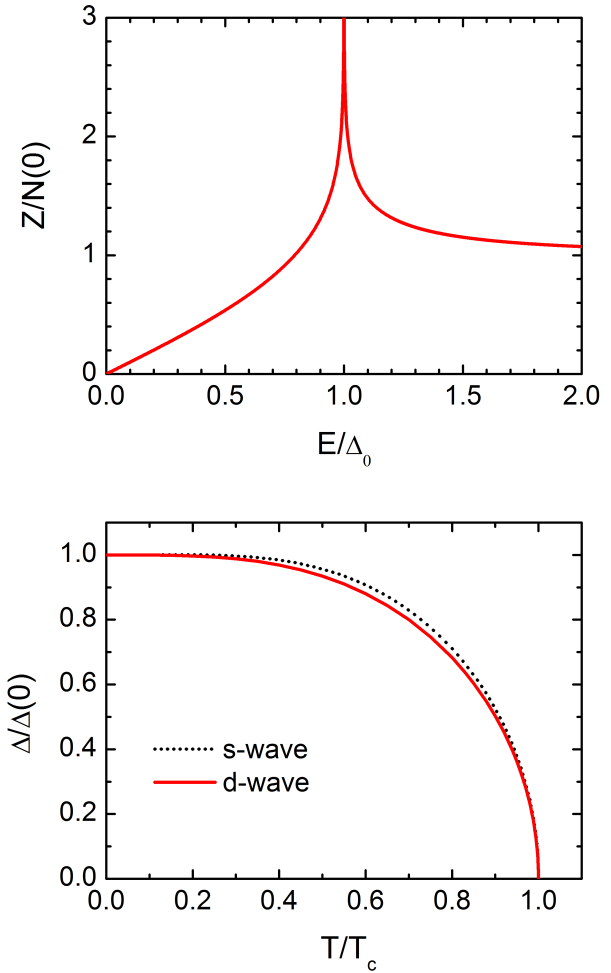

FIG. 3. Density of quasiparticle excitations and energy gap. The upper graph shows the density of quasiparticle excitations in the superconducting state (weak-coupling $d$-wave) normalized to the density of states at the Fermi edge in the normal-conducting state as function of the excitation energy normalized to the maximum value of the energy gap. In the lower graph the normalized energy gap as function of temperature (normalized to $T_c$) is shown for the weak-coupling $d$-wave (red curve) and $s$-wave (black dotted curve) case.

lower in the close vicinity of $E - E_F = 0$.

## IV. ANALYSIS OF SPECIFIC HEAT MEASUREMENTS

A method for calculating the specific heat of an unconventional superconductor from first principles has not been developed yet. Therefore, we have performed calculations of the specific heat of $\kappa$-Br in the framework of the multi-band alpha model for $d$-wave superconductors in order to check for compatibility of the gap parameters, as derived from our STS data, with specific

heat results from the literature[5]. The alpha model for $s$-wave superconductors, introduced by Padamsee et al.[6], extends BCS theory in a purely phenomenological way to materials with arbitrary (positive) values of the ratio $\alpha = \Delta(0)/k_B T_c$. As low temperature specific heat data for $\kappa$-Br indicate nodes in the gap function[5] ($d$-wave) as do our STS results (extended $s + d_{x^2-y^2}$), we extend the alpha model to the $d$-wave case. This can be done once the density of quasiparticle excitations $Z_d(E)$ and the gap function $\Delta_d(k,T)$ are known for this case. These two quantities have been calculated using BCS theory with a cylindrical Fermi surface and a gap function of the type $\Delta_d = \Delta_0(T)\cos(2\varphi)$. As the electronic system of $\kappa$-Br shows a strongly two-dimensional character, these approximations can be considered as an appropriate simplification for estimating the specific heat. The results for $Z_d(E)$ and $\Delta_0(T)$ are shown in Fig. 3. A value of $\alpha = 2.14$ has been obtained for the weak coupling limit consistent with the results of Ref. 7.

The multi-band alpha model, where each of the gaps occurs in a different band, deviates from the one-band scenario (described in the subsequent theoretical parts) which is used for fitting the STS data. However, as the specific heat depends essentially only on the quasiparticle density of states $Z(E)$, we attempt to fit this quantity (known from the theoretical considerations and the fitting of the STS data) by a sum of $d$-wave quasiparticle densities of states, each with its own maximum gap value $\Delta_{0,i}(T)$, i.e., with a multi-band alpha model. As an approximation to the density of states derived by the microscopic theory (blue line in Fig. 4) we use the function

$$[Z(E)/N(0)]_{\text{fit}} = \sum_{i=1}^{n} g_i\, z_d(E/\Delta_{0,i}(T)) \qquad (1)$$

with $z_d(E/\Delta_{0,i}(T))$ and $\Delta_{0,i}(T)/\Delta_{0,i}(0)$ being the $d$-wave functions shown in Fig. 3, $N(0)$ the density of states at the Fermi edge in the normal-conducting state and $\sum_{i=1}^{n} g_i = 1$. Here $g_i$ and $\Delta_{0,i}(0)$ are used as adjustable parameters. In our case three bands are needed, as the quasiparticle density of states shows three local maxima (see Fig. 4). The so-derived fitting function $[Z(E)/N(0)]_{\text{fit}}$ with the parameter values $g_1 = 0.08$, $g_2 = 0.54$, $g_3 = 0.38$ and $\Delta_{0,1}(0) = 0.48\,\text{meV}$, $\Delta_{0,3}(0) = 9.20\,\text{meV}$ is displayed in Fig. 4 (red dotted curve) and shows a fairly good agreement with the theoretical results. Using a $T_c$ value of 11.2 K, as obtained from the theoretical fit to the STS derived $\Delta_0(T)$ points, we get $\alpha_1 = 0.50$, $\alpha_2 = 2.67$, $\alpha_1 = 9.53$. The specific heat of the three-band alpha model is then given by

$$C = \sum_{i=1}^{3} C_d(T, T_c, \gamma_i, \alpha_i) \qquad (2)$$

with $\gamma_i = g_i\,\gamma$. The result of this calculation for the specific heat is shown in Fig. 5 (blue curve) together with

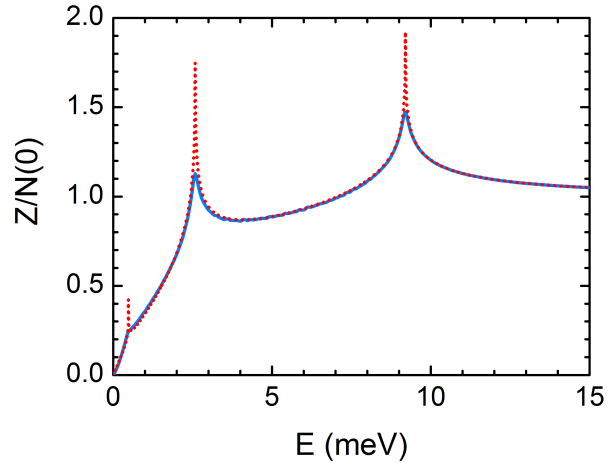

FIG. 4. Density of quasiparticle excitations for $\kappa$-Br. Normalized quasiparticle density of states for the parameter values (at $T = 5$ K) given in Table I of the main text with a small broadening of $\Gamma = 0.07\,\text{meV}$ (blue curve) and for the three-band alpha model (red dotted curve) with $g_1 = 0.08$, $g_2 = 0.54$, $g_3 = 0.38$ and $\Delta_{0,1}(0) = 0.48\,\text{meV}$, $\Delta_{0,2}(0) = 2.58\,\text{meV}$, $\Delta_{0,3}(0) = 9.20\,\text{meV}$.

the literature data from Ref. 5. The $\gamma$ value was set to $27.5\,\text{mJ}/(\text{mol K}^2)$ as determined by Ref. 5 and a Gaussian $T_c$ distribution ($\sigma = 0.98\,\text{K}$, $\overline{T}_c = 12.0\,\text{K}$) has been used to take into account the rounded maximum of the experimental data. The calculated results are in fairly good agreement with the experimental data. A better (and nearly perfect) fit can be obtained (red curve) by changing the $g_i$ values to $g_1 = 0.04$, $g_2 = 0.76$, $g_3 = 0.20$ and setting $\alpha_2 = 3.10$ (accounting for the dominant contribution to the specific heat), $\sigma = 0.83\,\text{K}$, $\overline{T}_c = 12.1\,\text{K}$.

## V. THEORY: BCS DENSITY OF STATES

The basic idea of BCS theory is the Cooper problem, which states that an attractive interaction between two electrons with opposite spin and momentum leads to a bound state, no matter how small it is. The corresponding Hamiltonian for Cooper pairs having a net momentum of zero can be written as

$$H = \sum_{k,\sigma} \epsilon_{k\sigma} c_{k\sigma}^\dagger c_{k\sigma} + \sum_{k,k'} U(k,k') c_{k\uparrow}^\dagger c_{-k\downarrow}^\dagger c_{-k'\downarrow} c_{k'\uparrow}.$$

The interaction can be treated in mean field theory ($\delta(c^\dagger c^\dagger) = c^\dagger c^\dagger - \langle c^\dagger c^\dagger \rangle$), where terms quadratic in $\delta$ are neglected. The resulting Hamiltonian can be diagonalized using the Bogoliubov-Valatin transformation, which introduces quasiparticle creation and annihilation operators $\gamma_{k\sigma}^\dagger$ and $\gamma_{k\sigma}$. The quasiparticle excitation energies are given as $E_k = \sqrt{\epsilon_k^2 + |\Delta_k|^2}$, where $\Delta(k) = \sum_{k'} U(k,k') \langle c_{-k'\downarrow} c_{k'\uparrow} \rangle$.

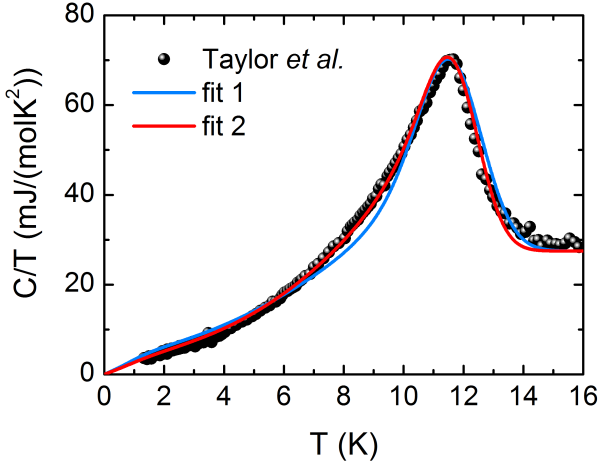

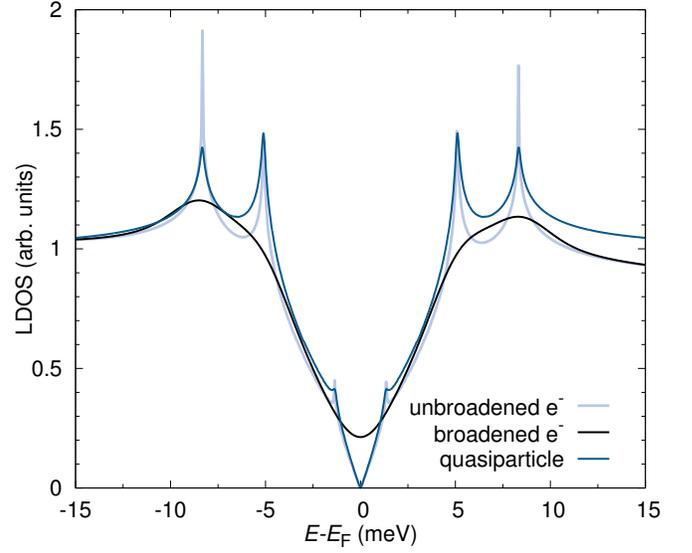

FIG. 5. Specific heat (devided by temperature). Black spheres are data from Taylor et al.[5], blue and red lines indicate our fits with the three-band alpha model for $d$-wave superconductors and a gaussian $T_c$ distribution. The parameters of fit 1 (fit 2) are: $\gamma = 27.5\,\mathrm{mJ/(mol\,K^2)}$, $g_1 = 0.08\,(0.04)$, $g_2 = 0.54\,(0.76)$, $g_3 = 0.38\,(0.20)$, $\alpha_1 = 0.50$, $\alpha_2 = 2.67\,(3.10)$, $\alpha_3 = 9.65$ and $\overline{T}_c = 12.0\,\mathrm{K}$ (12.1 K), $\sigma = 0.98\,\mathrm{K}$ (0.83 K).

FIG. 6. Comparison of broadened and unbroadened electron DOS to the quasiparticle DOS for identical parameters (as given in Table I). The quasiparticle DOS includes a small broadening of $\Gamma = 0.07$ meV. The energy scale is set to $\Delta_0 = 10$ meV.

The BCS Hamiltonian in terms of the quasiparticle creation and annihilation operators reads

$$H_{\mathrm{BCS}} = \sum_{k,\sigma} E_k \gamma_{k\sigma}^{\dagger} \gamma_{k\sigma} + \sum_k \epsilon_k$$
$$- \sum_{k,k'} U(k,k') \langle c_{k\uparrow}^{\dagger} c_{-k\downarrow}^{\dagger} \rangle \langle c_{-k'\downarrow} c_{k'\uparrow} \rangle.$$

The excitation spectrum of the quasiparticles $E_k$ is gapped and defined only for positive energies. The density of states of quasiparticles in an isotropic s-wave superconductor ($\Delta_k = \Delta_0$) can be calculated via

$$\rho_{\mathrm{qp}}(E) = \frac{1}{N} \sum_k \delta(E - E_k)$$
$$= \int d\epsilon\, \rho_0(\epsilon) \frac{\sqrt{\epsilon^2 + |\Delta|^2}}{\epsilon} \delta(\epsilon - \sqrt{E^2 - |\Delta|^2})$$
$$= \begin{cases} \rho_0\left(\sqrt{E^2 - |\Delta|^2}\right) \frac{E}{\sqrt{E^2 - |\Delta|^2}} & E > |\Delta| \\ 0 & E < |\Delta| \end{cases}.$$

As the quasiparticles are superpositions of particles and holes, this is not the density of states that can be measured in, for instance, scanning tunneling spectroscopy. To determine the electron DOS, we have to start from another commonly used expression of the density of states in terms of Greens functions, $\rho(E) = -\frac{1}{\pi} \sum_{k\sigma} \mathrm{Im}(-\langle c_{k\sigma}^{\dagger} c_{k\sigma} \rangle_0)$.

We assume a non spin-polarized energy dispersion and

again insert the Bogoliubov-Valatin transformation

$$\rho(E) = \frac{2}{\pi} \sum_k \mathrm{Im}\left( |u_k|^2 \left\langle \gamma_{k\uparrow}^{\dagger} \gamma_{k\uparrow} \right\rangle_0 + |v_k|^2 \left\langle \gamma_{-k\downarrow} \gamma_{-k\downarrow}^{\dagger} \right\rangle_0 \right).$$

As the Hamiltonian is diagonal in the quasiparticle operators, we can insert the expression for the bare Greens function and get

$$\rho(E) = 2 \sum_k \left( |u_k|^2 \delta(E - E_k) + |v_k|^2 \delta(E + E_k) \right),$$

where $k$ is a combined momentum and band index. $|u_k|^2$ and $|v_k|^2$ are the probabilities for the excitations being hole- or electron-like respectively. Only positive energies are taken into account[8].

Next we calculate the DOS explicitly for the *ab initio* derived bandstructure and the superconducting gap calculated microscopically. The delta functions in the DOS are replaced by gaussians with a standard deviation of 1.2 meV to simulate the broadening observed in experiment. For comparison we also calculated the DOS with a tetrahedron method[9] that does not employ any broadening. For the calculation of the DOS using gaussians we included $4000 \times 4000$ k-points in the xy-plane. In the tetrahedron method we used $4000 \times 4000 \times 2$ k-points.

A comparison of the broadened and unbroadened electron DOS and the quasiparticle DOS is shown in Fig. 6. Peak positions are identical. Only the background far away from the Fermi level is not well represented by the quasiparticle approximation. This can be easily understood from the structure of the expression derived before. To one energy $E$ two different electron energies $\epsilon_k$ and

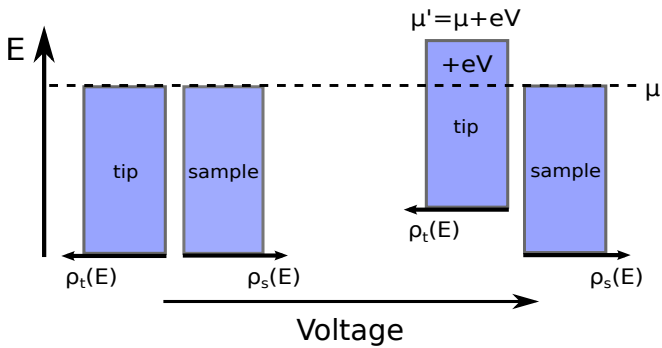

FIG. 7. Illustration of the energy levels in tip and sample with (right) and without (left) applied voltage. The voltage raises the energy levels of the tip, so that the chemical potential is higher than in the sample and a net current results. Note, that these energy shifts in the tip can be described in two ways: $f(E, \mu' = \mu + eV)$ or $f(E - eV, \mu)$.

$\epsilon_{k'}$ (where $\epsilon_k = -\epsilon_{k'}$) contribute and the BV coefficients should average to approximately $\sqrt{0.5}$. The small deviations are causing the asymmetric behavior, which merges to the normal density of states far away from the gap.

In the case of an anisotropic gap, the normal state density of states can not be identified easily, because the gap $\Delta$ then also depends on momentum $k$ and the Fermi surface is not a concentric circle.

$$\rho_s(E) = \int d\epsilon \frac{1}{N} \sum_k \delta(\epsilon - \epsilon_k) \delta(|E| - \sqrt{\epsilon^2 + |\Delta_k|^2})$$
$$\neq \int d\epsilon \rho_N(\epsilon) \delta(|E| - \sqrt{\epsilon^2 + |\Delta_k|^2}).$$

However, in the commonly used ansatz the electrons are considered to be free (Fermi surface is a concentric circle) and the gap is only determined by the angle $\theta$

$$\rho_s(E) = \frac{1}{(2\pi)^2} m_e \text{Re} \int d\theta \frac{|E|}{\sqrt{E^2 - |\Delta(\theta)|^2}}.$$

For all further calculations we will use this expression for the quasiparticle DOS.

## VI. THEORY: DIFFERENTIAL CONDUCTANCE

In a scanning tunneling spectroscopy (STS) experiment the current *from the tip to the sample* should obey (see Fig. 7),

$$I_{ts}(V) = \frac{2\pi e}{\hbar} \int_{-\infty}^{\infty} dE |M_{ts}(E - eV)|^2 \rho_{\text{tip}}(E - eV)$$
$$\times f(E - eV) \rho_{\text{sample}}(E)(1 - f(E)),$$

where $e$ is the positive elementary charge and a positive voltage $V$ corresponds to net charge carrier transport into the sample. The current is proportional to

the tunneling probability $|M_{ts}(E)|^2$, which depends on tip and sample geometry. Furthermore, the tunneling probability is proportional to the number of occupied states in the tip $\rho_{\text{tip}}(E - eV) f(E - eV)$ and the number of unoccupied states in the sample at lower energies $\rho_{\text{sample}}(E)(1 - f(E))$.

After shifting the integration variable $E \to E + eV$, we arrive at the literature form of the expression for the tunneling current. In order to obtain the net current, we have to calculate the difference between the currents from *tip to sample* and *sample to tip*. We obtain

$$I(V) = \frac{2\pi e}{\hbar} \int_{-\infty}^{\infty} dE |M_{ts}(E)|^2 \rho_{\text{tip}}(E) \rho_{\text{sample}}(E + eV)$$
$$\times (f(E) - f(E + eV)).$$

The tunneling matrix element $|M_{ts}(E)|^2$ is experimentally not accessible and therefore the tunneling current is divided by the voltage dependent tunneling transmission function $T(V)$, which can be determined from the experimental setup. The density of states of the tip is considered to be independent of energy in the region of interest. We obtain the differential conductance (shifting once again the integration variable $E \to E - eV$) by taking the derivative with respect to the voltage $V$,

$$\frac{1}{T(V)} \frac{dI(V)}{dV}$$
$$= \frac{2\pi e}{\hbar} \rho_{\text{tip}}(E_F) \times \int_{-\infty}^{\infty} dE \rho_{\text{sample}}(E) \left( -\frac{df(E + eV)}{dV} \right).$$

## VII. THEORY: BROADENING

In the preceding expression we already included thermal broadening via the Fermi function. Additionally we have to take into account the finite quasiparticle lifetime. This can be done (following Dynes *et al.*[10]) by replacing the energies $E$ by $E + i\Gamma$ in the density of states. If we now consider only proportionalities, we can also insert the angle dependent expression for the sample density of states

$$\frac{dI(V)}{dV} \propto T(V) \int_{-\infty}^{\infty} dE \int_0^{2\pi} d\theta$$
$$\times \text{Re} \frac{|E + i\Gamma|}{\sqrt{(E + i\Gamma)^2 - \Delta(\theta)^2}} \left( -\frac{df(E + eV)}{dV} \right).$$

Note that this formula can not be used to fit data with large finite values of the differential conductance at zero bias voltage, since the superconducting DOS goes to zero for zero bias voltage, independent of the choice of $\Gamma$.

Sometimes a slightly different expression is used for the Dynes broadening: $|\text{Re}[(E + i\Gamma)/\sqrt{(E + i\Gamma)^2 - \Delta(\theta)^2}]|$ Although only taking the real part and calculating the absolute value were exchanged here, this formula yields a finite value for the superconducting DOS at zero bias voltage, which is proportional to $\Gamma$. Therefore, one can

| name | function | coefficient | value |
|------|----------|-------------|-------|
| $s\pm$ | $\cos k_x + \cos k_y$ | $c_{s_1}$ | 0.069 |
| $s\pm$ | $\cos k_x \cdot \cos k_y$ | $c_{s_2}$ | -0.672 |
| $d_{x^2-y^2}$ | $\cos k_x - \cos k_y$ | $c_{d_1}$ | -0.259 |
| $d_{xy}$ | $\sin k_x \cdot \sin k_y$ | $c_{d_2}$ | 0.000 |

TABLE I. Basis functions considered in the fitting process of the microscopic result for the gap function. Given values are the result of the fitting procedure.

use it to fit data for the differential conductance, which include a constant background. Unfortunately, exchanging the order of the real part and absolute value operations is inconsistent with the expression for the BCS density of states. We fix this problem by introducing the shift of the background explicitly.

## VIII. THEORY: FITTING GAP SYMMETRIES TO STS DATA

On top of the broadening effects, it was found that the measured density of states is suppressed by the formation of a soft Hubbard gap caused by short-range interactions and disorder[2]. This correction can be taken into account by scaling the density of states by $B(V)$ as described before. Including this correction the differential conductance reads

$$\frac{1}{B(V)T(V)}\frac{dI(V)}{dV} \propto \int_{-\infty}^{\infty} dE \int_{0}^{2\pi} d\theta$$
$$\times \operatorname{Re} \frac{|E+i\Gamma|}{\sqrt{(E+i\Gamma)^2 - \Delta(\theta)^2}}$$
$$\times \left(-\frac{df(E+eV)}{dV}\right).$$

For anisotropic gaps on Fermi surfaces, which are not concentric circles, this equation is still an approximation. To improve our calculations, we execute the integration over the angle as a summation over points on the discretized two-dimensional *ab initio* Fermi surface. We find from our microscopic calculations that the symmetry of the superconducting gap can be described by a superposition of two $s\pm$ and the $d_{x^2-y^2}$ functions, $\Delta(k_x, k_y) = \Delta_0[(c_{s_1}(\cos(k_x) + \cos(k_y)) + c_{s_2}\cos(k_x)\cos(k_y) + c_{d_1}(\cos(k_x) - \cos(k_y))]$, where the normalized prefactors $c_i$ ($\sum_i |c_i| = 1$) can be a) fitted to the experiment (see Fig. 8), *or* b) set to the microscopic values (see Fig. 9), so that only one common prefactor $\Delta_0$ is fitted to the magnitude of the gap.

The Dynes broadening $\Gamma$ is used as a fit parameter. The value for the differential conductance in the measured data after dividing by correction terms $B(V)$ and $T(V)$ is still finite. This finite differential conductance at zero bias voltage is too large to be explained by thermal broadening. Therefore, we introduce a parameter

$x$, which scales and shifts the calculated DOS so that a constant background in the differential conductance is subtracted, while keeping the data points far away from the Fermi level at unity. With this additional correction we obtain the final formula given in the main text.

The gap on the Fermi surface is shown in Fig. 10 for both the microscopic ratios and those obtained from the fit to the STS data. The analytic representation for the microscopic results was determined by fitting a linear combination of symmetry functions given in Table I. The largest deviation of the fit from the microscopic gap per evaluated k-point is 4.55% of the maximum gap value. The average deviation of the fit is 1.25% of the maximum gap value. For a comparison of the microscopic result to its analytic representation see Fig. 10.

While Fig. 10 emphasizes the qualitative agreement between experiment and theory, we include also polar and linear plots (see Fig. 11) of the same data to reveal some quantitative differences. The angle $\varphi$ in the $k_x$-$k_y$ plane is measured from the $k_x$ direction. For the RPA curves and the fitted result the energy scale was set to $\Delta_0 = 10$ meV and $\Delta_0 = 12.1$ meV respectively. In this way, the RPA result can be compared directly to the data shown in Fig. 6 or the DOS shown in Fig. 4 of the main paper, while the data obtained by analysing the experiment can be compared to Fig. 3 in the main paper.

While the symmetry of the superconducting gap obtained from theory is identical to the one found by fitting the experiment, the size of the superconducting gap around $\varphi \approx 45°$ is clearly overestimated in our calculation. Quantitative differences are however to be expected, as our RPA approach does not take into account the electronic self-energy and lacks a self-consistency condition. Note that there is currently no state-of-the-art method that can quantitatively predict the momentum-resolved gap structure or superconducting transition temperature of two-dimensional correlated electron systems.

Finally, we investigate whether a nodeless anisotropic s-wave symmetry could be present in the system. We combine a plain s-wave gap with the d-wave expressions given in our manuscript and determine the prefactors for the contributions that lead to optimal agreement with the experimental spectrum measured at 5 K (see Fig. 12). For both the $s + d_{xy}$ and the $s + d_{x^2-y^2}$ case we observe that the d-wave component is dominant in the optimal fit to the STS data. This excludes a nodeless anisotropic s-wave symmetry. The agreement of the $s + d_{xy}$ symmetry is considerably worse than for the $s + d_{x^2-y^2}$ symmetry. The $s + d_{x^2-y^2}$ and the $s\pm + d_{x^2-y^2}$ solution presented in our manuscript describe the data equally well. However, as a plain s-wave component cannot be caused by antiferromagnetic spin-fluctuations alone, we decided to use the physically more meaningful representation in terms of an extended s-wave component.

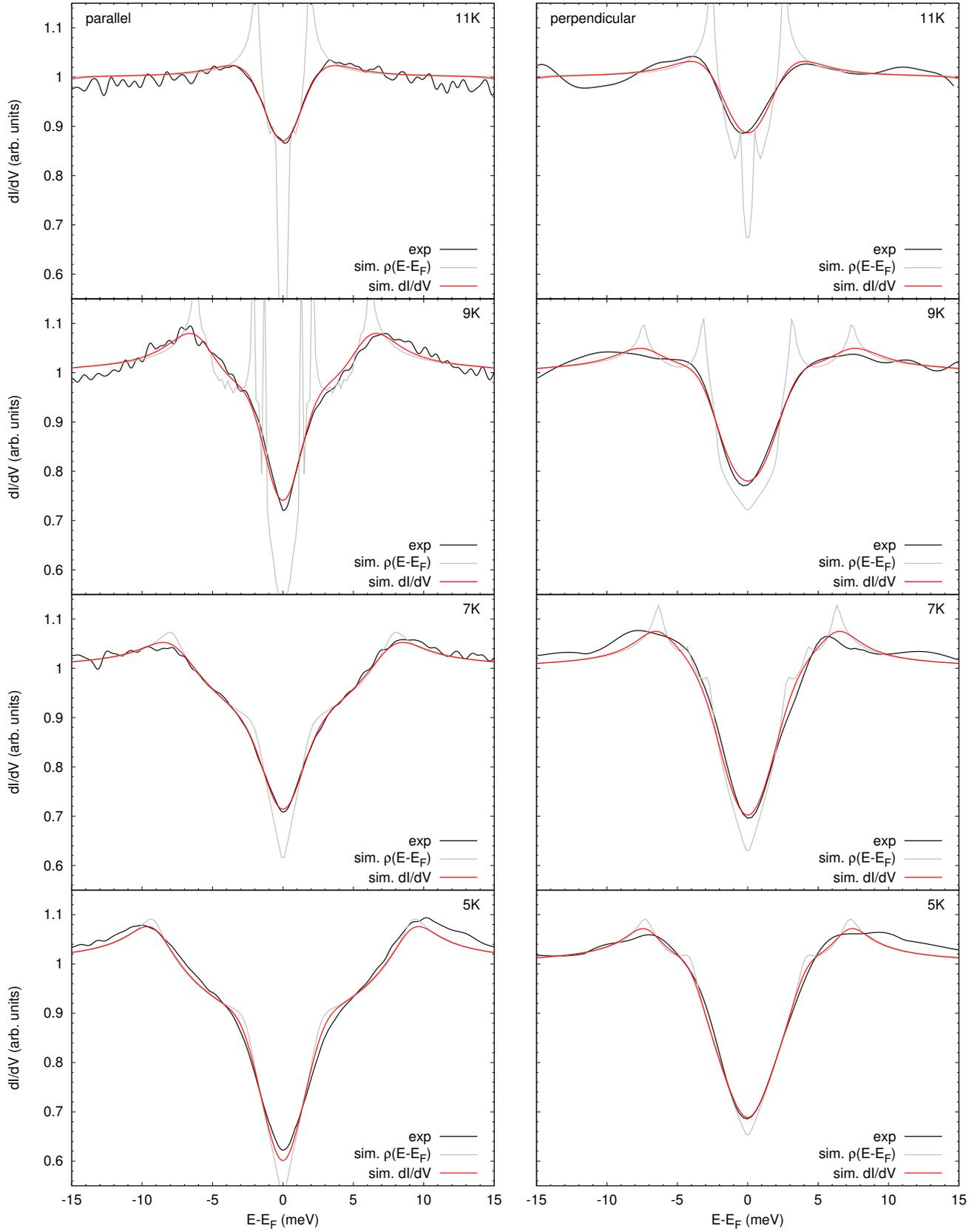

FIG. 8. Fit to the experimental data at the different temperatures (Prefactors are fitted).

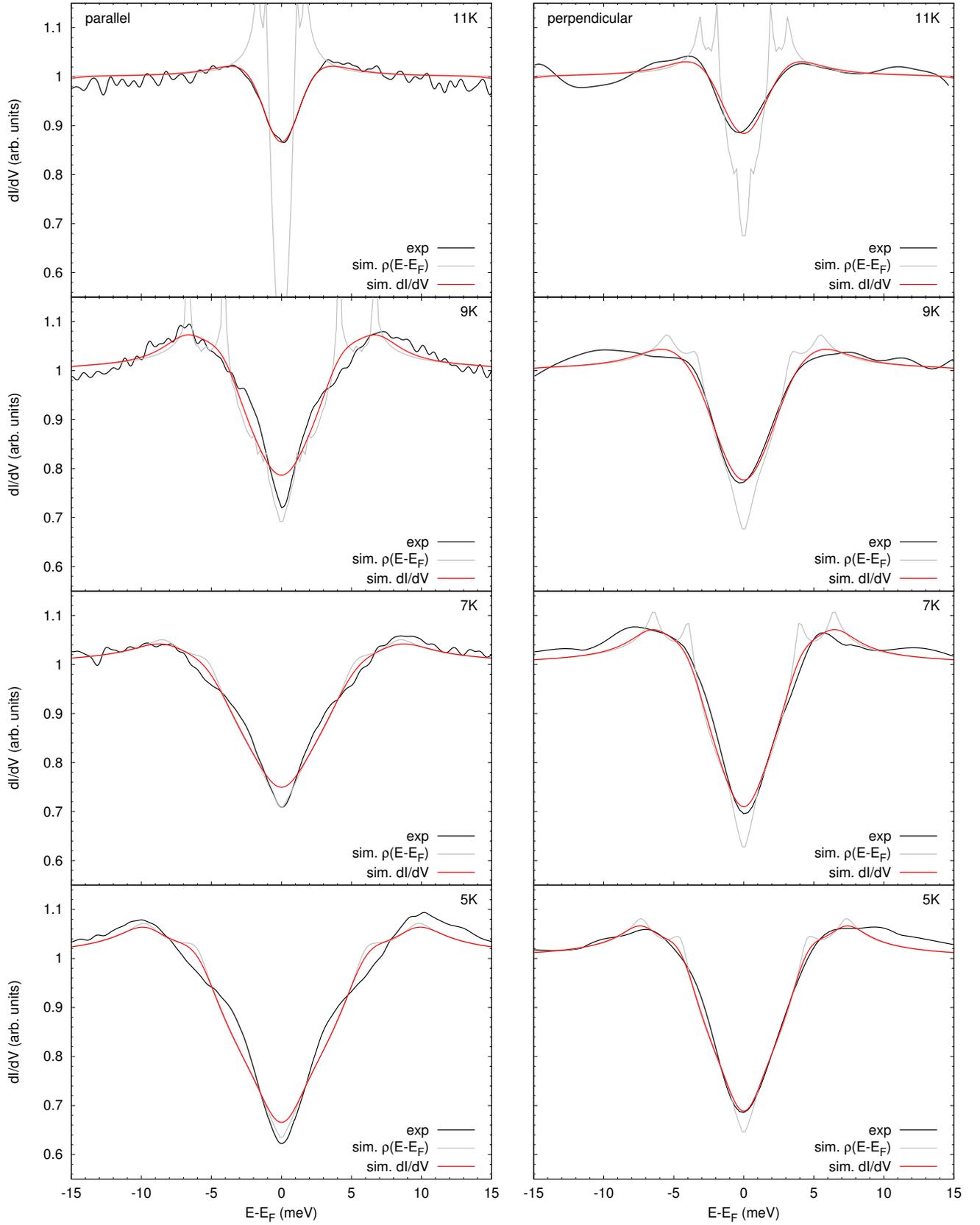

FIG. 9. Fit to the experimental data at the different temperatures (Prefactors fixed to microscopic values).

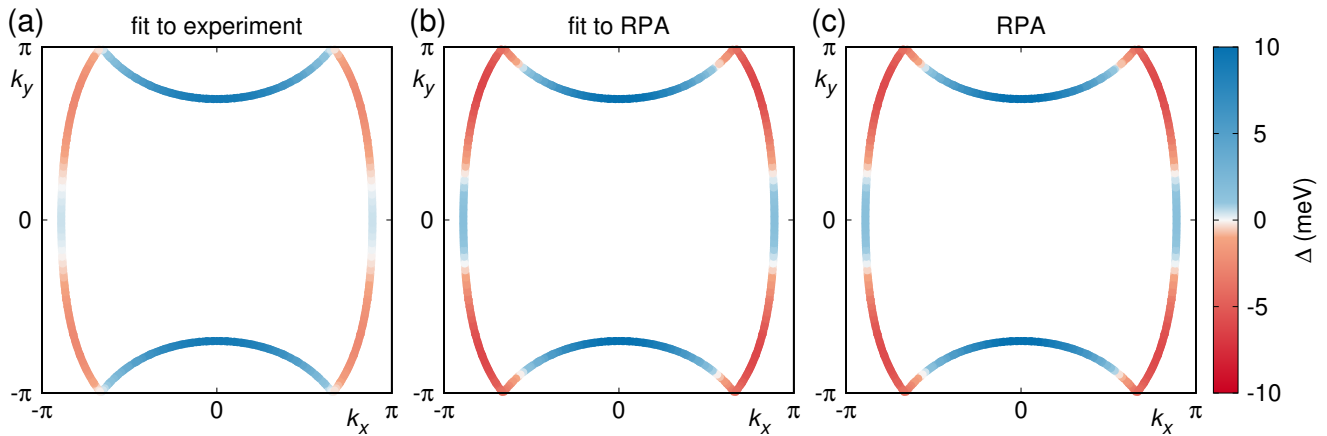

FIG. 10. Comparison of the gap functions on the *ab initio* Fermi surface obtained from (a) fitting the STS experiment (b) fitting the microscopic result calculated using RPA and (c) the microscopic result itself. All data points of the symmetry functions were multiplied by the energy scale $\Delta_0 = 12.1$ eV.

## IX. THEORY: TEMPERATURE DEPENDENCE OF THE GAP

To estimate values for the critical temperature and the maximum gap size at zero temperature, we fitted our temperature dependent gap values to the interpolation formula $\Delta(T) = \Delta_0 \tanh(1.74\sqrt{\frac{T_c}{T} - 1})$ for the solution of the s-wave BCS self consistency equation (see Fig. 13).

Solving the BCS self-consistency equation for unconventional pairing mechanisms is difficult due to the momentum dependence of the pairing interaction. Therefore, we decided to use the s-wave solution as a rough approximation.

Our predicted values for the critical temperatures $T_c = 10.3 \pm 1.2$ K for fitted prefactors and $T_c = 11.2 \pm 0.2$ K for fixed prefactors are in good agreement with the experimental observation of $T_c \approx 11.5$ K (see f.i. Ref. 11). The maximal gap size is found to be $\Delta_0 = 12.9 \pm 2.0$ meV (fitted prefactors) and $\Delta_0 = 12.1 \pm 0.7$ meV (fixed prefactors).

## X. THEORY: AB-INITIO CALCULATIONS

We use the experimental crystal structure[12], but relax the ethylene endgroups of the BEDT-TTF molecules in eclipsed configuration[13]. For the exchange correlation functional we use the generalized gradient approximation (GGA)[14]. The DFT calculation was converged using $6 \times 6 \times 6$ k-point grids.

## XI. THEORY: RPA SPIN-FLUCTUATION PAIRING, FORMALISM

In $\kappa$-(BEDT-TTF)$_2$X materials there is strong evidence for antiferromagnetic spin-fluctuations[15]. There-fore, we investigate the superconducting state of these materials based on a random phase approximation (RPA) spin-fluctuation approach[16]. We have generalized our implementation from single-site multi-orbital models[17,18] to multi-site single-orbital models relevant for the materials discussed here.

The low-energy Hamiltonian is given by the kinetic part $H_0$, derived with the Wannier function method described in the main text, and the intra-orbital Hubbard interaction $H_{int}$.

$$H = H_0 + H_{int}$$
$$= \sum_{\sigma} \sum_{\langle ij \rangle} t_{ij} c_{i\sigma}^{\dagger} c_{j\sigma} + \frac{U}{2} \sum_{\sigma} \sum_i n_{i\sigma} n_{i\bar{\sigma}}$$

Here, $\sigma$ represents the spin and $n_{i\sigma} = c_{i\sigma}^{\dagger} c_{i\sigma}$. The sum over $i$ runs over all BEDT-TTF sites in the unit cell. The hopping integrals $t_{ij}$ taken into account are not restricted to nearest or next-nearest neighbor hoppings. We rather determine the optimal distance cutoff during the wannierization procedure. The interaction strength $U$ is treated as a parameter.

We calculate the non-interacting static susceptibility $\chi^0$, where matrix elements $a_\mu^l(\vec{k})$ resulting from the diagonalization of the initial Hamiltonian $H_0$ connect orbital and band-space denoted by indices $l$ and $\mu$ respectively. The $E_\mu$ are the eigenvalues of $H_0$ and $f(E)$ is the Fermi function.

$$\chi^0_{spqt}(\vec{q}) = -\frac{1}{N_k} \sum_{\vec{k},\mu,\nu} a_\mu^s(\vec{k}) a_\mu^{p*}(\vec{k}) a_\nu^q(\vec{k}+\vec{q}) a_\nu^{t*}(\vec{k}+\vec{q})$$
$$\times \frac{f(E_\nu(\vec{k}+\vec{q})) - f(E_\mu(\vec{k}))}{E_\nu(\vec{k}+\vec{q}) - E_\mu(\vec{k})}$$

In our calculation both $\vec{q}$ and $\vec{k}$ run over uniform grids spanning the reciprocal unit cell. Temperature enters the calculation through the Fermi functions.

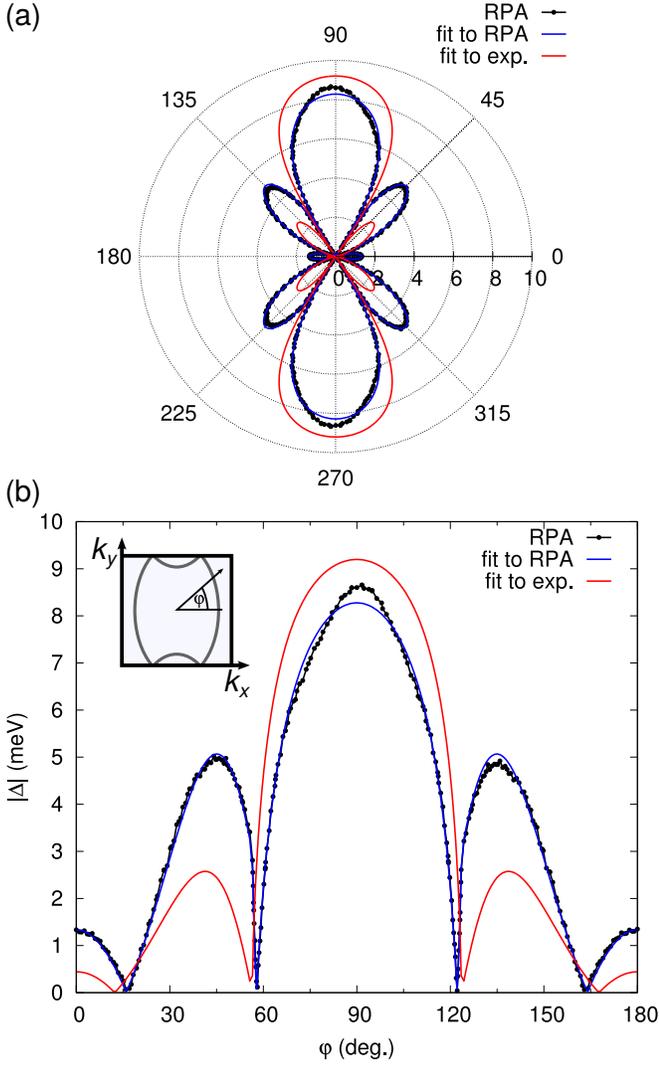

FIG. 11. (a) Polar and (b) linear plot of the magnitude of the superconducting gap $|\Delta|$ on the Fermi surface versus the angle $\varphi$ measured from the $k_x$ direction. For the RPA curves and the fit to experiment the energy scale was set to $\Delta_0 = 10$ meV and $\Delta_0 = 12.1$ meV respectively.

The static spin- and orbital-susceptibilities ($\chi^{s,\mathrm{RPA}}$ and $\chi^{c,\mathrm{RPA}}$) are constructed in an RPA framework. Since the interaction term in the Hamiltonian is local and our models are single-orbital in nature, we can restrict the calculation to the diagonal elements of the susceptibility and use scalar equations for the RPA-enhanced susceptibilities.

$$\chi_L^{s,\mathrm{RPA}}(\vec{q}) = \frac{\chi_L^0(\vec{q})}{1 - U\chi_L^0(\vec{q})}, \quad \chi_L^{c,\mathrm{RPA}}(\vec{q}) = \frac{\chi_L^0(\vec{q})}{1 + U\chi_L^0(\vec{q})}$$

Here, $\chi_L$ with $L = \{llll\}$ denotes the diagonal element of the susceptibility tensor associated with an BEDT-TTF site indexed by $l$. The total spin susceptibility is given by the sum over all site-resolved contributions $\chi^s(\vec{q}) = \frac{1}{2}\sum_L \chi_L^{s,\mathrm{RPA}}(\vec{q})$.

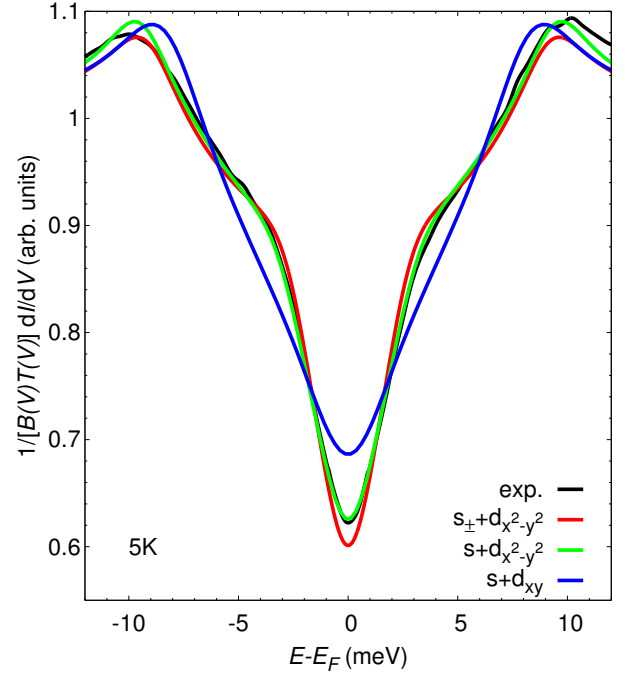

FIG. 12. Experimental spectrum at 5 K and simulation presented in the main paper ($s_\pm + d_{x^2-y^2}$) compared to simulated spectra of plain $s$-wave combined with $d_{xy}$ and $d_{x^2-y^2}$ expressions.

The pairing vertex in orbital space for the singlet channel can be calculated using the fluctuation exchange approximation[19,20].

$$\Gamma_{spqt}(\vec{k}, \vec{k}') = [\frac{1}{2}U\chi^{s,\mathrm{RPA}}(\vec{k} - \vec{k}')U + U\chi^{s,\mathrm{RPA}}(\vec{k} + \vec{k}')U$$
$$-\frac{1}{2}U\chi^{c,\mathrm{RPA}}(\vec{k} - \vec{k}')U + U]_{spqt}$$

Momenta $\vec{k}$ and $\vec{k}'$ are restricted to the Fermi surface. As vectors $\vec{k} \pm \vec{k}'$ do not necessarily lie on the grid used in the calculation of the susceptibility $\chi^0(\vec{q})$, we use linear interpolation of the grid data.

The pairing vertex in orbital space is transformed into band space using the matrix elements $a_\mu^l(\vec{k})$.

$$\tilde{\Gamma}_{\mu\nu}(\vec{k}, \vec{k}') = \mathrm{Re}\sum_{spqt} a_\mu^{t,*}(\vec{k})a_\mu^{p,*}(-\vec{k})[\Gamma_{spqt}(\vec{k}, \vec{k}')]$$
$$\times a_\nu^s(\vec{k}')a_\nu^q(-\vec{k}')$$

Finally, we solve the linearized gap equation by performing an eigendecomposition on the kernel and obtain the dimensionless pairing strength $\lambda_i$ and the symmetry function $g_i(\vec{k})$.

$$-\sum_\nu \oint_{C_\nu} \frac{dk'_\parallel}{2\pi} \frac{1}{2\pi\, v_F(\vec{k}')} \left[\tilde{\Gamma}_{\mu\nu}(\vec{k}, \vec{k}')\right] g_i(\vec{k}') = \lambda_i g_i(\vec{k})$$

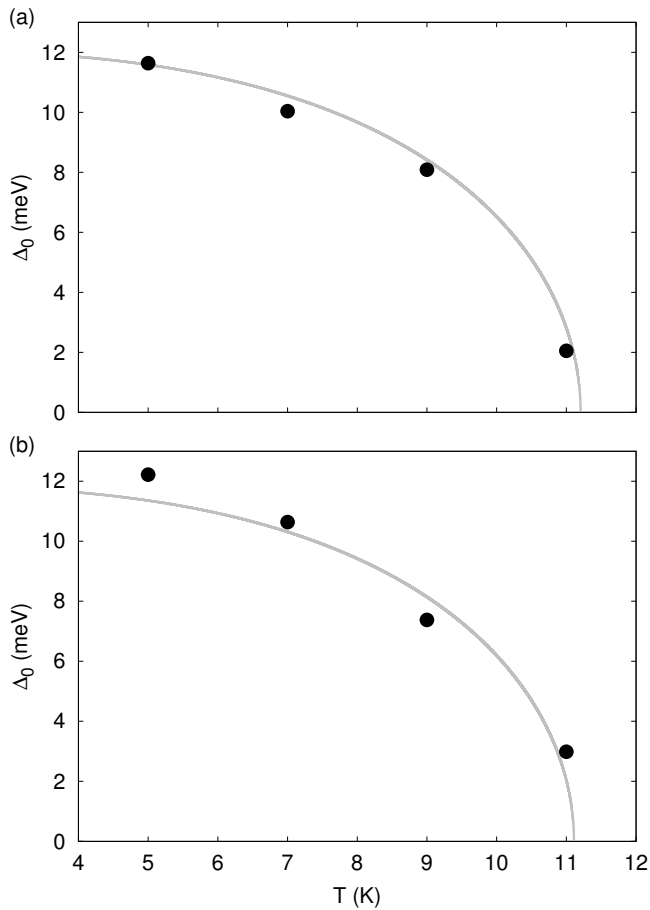

FIG. 13. Fit of the interpolation function for the solution of the s-wave BCS self consistency equation to the temperature dependent gap values obtained from the STS analysis. (a) shows the gaps extracted from the fit using prefactors $c_i$ fixed to the values determined from the microscopic calculation, while (b) shows the gaps extracted from a fit to the STS experiment, where also the $c_i$ were used as parameters.

In the gap equation we use the singlet symmetrized vertex $\Gamma_{\mu\nu}(\vec{k}, \vec{k}') = \frac{1}{2}[\tilde{\Gamma}_{\mu\nu}(\vec{k}, \vec{k}') + \tilde{\Gamma}_{\mu\nu}(\vec{k}, -\vec{k}')]$. The integration runs over the discretized Fermi surface and $v_F(\vec{k})$ is the magnitude of the Fermi velocity.

## XII. THEORY: RPA SPIN-FLUCTUATION PAIRING, COMPUTATIONAL DETAILS

For the intra-molecular Coulomb interaction we use a value of $U = 0.75$ eV. The non-interacting susceptibility is calculated on a $50 \times 50$ k-point grid at an inverse temperature of $\beta = 40$ eV$^{-1}$. For the solution of the gap equation we used 548 points on the two-dimensional Fermi surface.


\end{document}